\documentstyle[12pt]{article}
\include{epsf}
\epsfverbosetrue
\def\thetaind{\theta^{ind}}
\def\la{\langle}
\def\ra{\rangle}
\newcommand{\beq}{\begin{equation}}
\newcommand{\eeq}{\end{equation}}
\newcommand{\bea}{\begin{eqnarray}}
\newcommand{\eea}{\end{eqnarray}}

\begin{document}
\begin{titlepage}

\begin{flushright}

hep-ph/0006057
\end{flushright}
\vskip1.8cm
\begin{center}
{\LARGE
Induced $\theta$-Vacuum States in Heavy Ion Collisions:\\       
\vskip0.6cm
A Possible Signature.
}         
\vskip1.5cm
 {\Large Kirk Buckley ,}
{\Large Todd Fugleberg} 
and
{\Large Ariel Zhitnitsky}
\vskip0.5cm
        Physics and Astronomy Department \\
        University of British Columbia \\
 6224 Agricultural Road, Vancouver, BC V6T 1Z1, Canada \\ 
        {\small e-mail:
kbuckley@physics.ubc.ca \\ 
fugle@physics.ubc.ca \\
arz@physics.ubc.ca }\\
\vskip1.0cm
\vskip1.0cm
{\Large Abstract:\\}
\end{center}
\parbox[t]{\textwidth}{
It has been suggested recently \cite{KFZ} that  an arbitrary induced 
$\theta$-vacuum state ($\theta^{ind}$)
could be created in heavy ion collisions. If such a
state can be created, it would decay by various mechanisms to the
$\theta^{fund}=0$ state which is the true ground state of  our world. 
In the following we
will discuss the possibility of studying this unusual state
through the emission of pions, $\eta$-mesons, and $\eta'$-mesons. 
We will also present the spectrum of the
produced particles in this non-zero $\theta$ background. 
We use the instantaneous perturbation theory for our estimates.}

\vspace{1.0cm}

\end{titlepage}

\section{Introduction}

The long awaited Relativistic Heavy Ion Collider at 
Brookhaven (RHIC) could yield many fascinating results concerning
the accepted theory of the strong force, QCD. Although the main 
goal of this experiment is to study the quark-gluon plasma, there exist many 
other possibilities. One of the most sought after results is the
nature of the chiral 
symmetry restoration and/or deconfining phase transition.
Another possible outcome is the so called disoriented chiral
condensate (DCC). In addition, it was recently suggested \cite{KFZ} 
that it may be possible to create an induced $\theta \neq 0$-state 
in heavy ion colliders. If this is possible, the consequences could 
include the observation of a new state of matter.

Indeed, 
as is known, in the infinite volume limit
and in   thermal equilibrium the $\theta$-vacuum state
is the absolutely stable ground state  of a new world with 
new physics  
quite  different from ours. 
In particular, $P$ and $CP$ symmetries are
strongly violated in this world. In spite of the fact that the 
$\theta$-vacuum state has a higher energy (see below)
this state is stable 
due to the super selection rule:
There does not exist a gauge invariant observable  
$\cal{A}$ in QCD, which would communicate between 
two different worlds, 
$\la\theta'|\cal{A}|\theta\ra\sim \delta(\theta-\theta')$.
Therefore, there are no transitions between these worlds\cite{gross}.

Of course, we do not expect that such a stable
$\theta$-state can be produced
in heavy ion collisions. However, we do expect
that locally, for a short period of time
due to the non-equilibrium dynamics after the
QCD phase transition, a large domain with wrong 
$\theta$ direction may be formed.  
This provides us with a unique opportunity to study
a new state of matter.
The idea is very similar to the old idea of the 
creation of the Disoriented Chiral 
Condensate (DCC) in heavy ion collisions \cite{DCC} -\cite{Rajagopal}.
DCC refers to regions of space (interior) in which the chiral condensate points 
in a different direction from that of the ground state (exterior), and separated 
from the latter by a hot shell of debris. 
Our starting point is the conjecture
that a classically large domain with a nonzero
$U(1)_A$ chiral phase may be formed in a  heavy ion collisions, i.e. we assume
 that the expectation value for the chiral condensate in a sufficiently large 
domain has, in general,  a non-zero $U(1)$ phase:
$\la \bar{\Psi}_L \Psi_R \ra \sim
e^{i\phi^{singlet}}| \la \bar{\Psi} \Psi \ra | $.
This phase is identified with $\theta^{ind}$. Such an identification 
is a direct consequence of the transformation properties of the 
fundamental QCD Lagrangian under $U(1)_A$ rotations when the chiral 
singlet phase can be rotated away at the 
cost of the introduction of the induced $\theta$, 
$\theta^{ind}=-N_f\phi^{singlet}$. 

We should remind the reader at this point that
in QCD, the fundamental $\theta$ parameter enters in the combination
$\theta Q=\theta \frac{\alpha_s}{8 \pi} G^a \tilde{G^a}$. This gives 
us the first contribution to the physically observable
 parameter $\overline{\theta}$. 
The second contribution is related to the phase which is introduced
upon diagonalization of the quark mass matrix, which can be in 
principle arbitrary. 
Current experimental results indicate that 
these two numbers cancel with precision better than $10^{-9}$. 
The problem
of why $\overline{\theta}\simeq 0
$ in our world is known as the strong $CP$-problem, and we do not address
this problem in the present paper. Our point is the following:
by having a non-zero $U(1)$ phase
$\la \bar{\Psi}_L \Psi_R \ra \sim
e^{i\phi^{singlet}}| \la \bar{\Psi} \Psi \ra | $   
  in a sufficiently large 
region, we essentially break this precise cancellation between 
two contributions in a macroscopically large domain\footnote{We use the term
macroscopic in this paper to refer to scales which are larger than  
the fundamental scale, $\Lambda_{QCD}^{-1}$}. 
Exactly this non-cancellation will mimic a non-zero $\theta$-parameter.  

The production of non-trivial $\thetaind$-vacua
would occur in much the same way as discussed previously for  
DCC\cite{DCC}-\cite{Rajagopal}. The new element is that, in addition to 
chiral fields differing from their true vacuum values,
the induced $\theta$-parameter,
which is zero in the real world, becomes effectively nonvanishing in the
macroscopically large domain as explained above. Therefore, 
although the fundamental $\theta$-parameter remains zero, 
we would expect that an induced $\theta$-parameter
can be  non-zero in a macroscopically large domain, and it
will mimic the physics of the world with $\theta^{fund} \neq 0$.
A different name for the same object
would be a `` zero (spatially independent) mode  of the $\eta'$ field"
which exactly corresponds to the
  flavor-singlet spatially independent  part of the chiral condensate phase.
However, we prefer to use the term $\thetaind$
because  the symbol $\eta'$ is usually associated with the 
free asymptotic state of the heavy $\eta'$-meson 
(excitation) and not with a classical constant field (condensate) 
in a large domain where $\eta'$-mesons
interact very strongly and change their properties
in this background. This other terminology could be also misleading
because of this reason. In addition to the above mentioned difference, 
the disoriented chiral condensate involves the amplification of the low 
momentum modes of the charged pions, while the formation the $\theta$-state
involves the amplification of the low momentum modes of only the 
neutral particles, including the $\eta'$ singlet . 

One might ask, ``How could such a $\thetaind$-state be detected if it
is created in heavy ion collisions for a very short period of time?''.
One obvious  possibility is that the
$\thetaind$-state could be observed through Goldstone bosons with
specific $CP$-odd correlations \cite{Pisarski}.  
However, a more promising signature is related to 
  the low momentum modes of the chiral fields\cite {arz}. Indeed,
as was shown in \cite{KFZ}, the creation of
the $\thetaind$-state involves the enhancement of the low
frequency modes of the neutral chiral fields. 
Therefore, it is natural to expect that particles
made of these chiral fields are to be produced when the  
$\thetaind$-state blows apart.
To be more precise, in the limit $m_u=m_d=m_s\ll \Lambda_{QCD}$
the only relevant physical degree of freedom is a singlet phase
of the chiral condensate which transforms, at the very end of the transition
exclusively, to the $\eta'$-mesons when the  $\thetaind$-state
falls apart. Therefore, in this limit we would expect
an excess of $\eta'$-mesons to be produced.
 However, due to the mixing and difference in masses, $m_u\neq m_d\neq m_s $, 
it is expected that all neutral modes are to be produced.
 The main goal of the present work is
  to study the spectrum and the angular distribution
of particles which will be produced when the  $\thetaind$-state
falls apart.

 However, before we discuss some numerical estimates, 
we would like to present 
  some estimates based on simple
energetic considerations. As we discuss  below (\ref{V_vac}),
the vacuum energy density in the $\theta$-state is greater than in the
$\theta=0$ vacuum state 
by the amount: $$\Delta E=
E_{vac}(\theta)- E_{vac}(\theta=0)=2m|\la \bar{q} q \ra |
(1-|\cos\frac{\theta}{2}|) ,$$
where for simplicity we use $ N_f=2 $ with $m_u=m_d=m$.
When the $\thetaind$-state blows apart, the  energy
associated with this background will be released as 
Goldstone bosons carrying the total energy:
$$
\Delta E ~V \simeq 2m|\la \bar{q} q \ra |
(1-|\cos\frac{\theta}{2}|)\cdot V\simeq 20 
\cdot\left(\frac{V}{(10~fm)^3}\right)~GeV ,
$$
where $V=L^3$ is $3d$ volume of the $\thetaind$-background
measured in $Fermi$.
Therefore, given the $cm$ energy $\sqrt{s}=40~TeV$, only
a small fraction: 
$$\rho\sim \frac{\Delta E V}{\sqrt{s}}\sim \frac{ 20 
 (\frac{V}{(10~fm)^3})~GeV}{40~TeV}\sim 10^{-3} 
\left( \frac{V}{(10~fm)^3} \right) $$
of the total collision energy will be released through
the decay   of the induced $\thetaind$-state.
Two of the most profound features of the Goldstone bosons produced 
in this decay are the following: \\ 
1. Due to the fact that the 
 $\thetaind$ formation is caused by amplification of low 
momentum modes, we expect the spectrum of 
the Goldstone bosons from this decay to
be  strongly enhanced at low $|\vec{k}|\sim L^{-1}$;\\
 2.
Due to the nonzero value for the topological density
$\langle \theta | \frac{\alpha_{s}}{8 \pi } G \tilde{G}
| \theta \rangle =-\frac{\partial V_{vac}(\theta)}{\partial\theta}=
-m|\la 0| \bar{q} q |0 \ra |\sin\frac{\theta}{2}$, see
(\ref{V_vac}), one could expect that some $P,~CP$ odd correlations
would appear in this background.  

Considering this, we should look for
enhanced production of particles such as the $\pi,\eta,$ and $\eta'$ with low
momentum, on the  $(10-100)~MeV $
scale (depending on the size of the domain
$L$). The difference between the $\theta$-vacuum state and the DCC which
also involves enhancement of low momentum modes is that the $\theta$-vacuum
state also produces $\eta'$-mesons.
These should decay eventually by various processes to
photons and dileptons, and if a large number of these low momentum particles 
are detected this would be a definite signal of
the creation of a macroscopically large domain
with $\thetaind \neq 0$. One could 
further speculate that such an enhancement  could also account
for the unexplained large number of low momentum dileptons seen at
CERN \cite{cern} as suggested   in\cite{arz}. However, detailed calculations 
are needed to give this speculative conjecture further support. 

The presentation is organized as follows. In Sect. 2 we briefly describe the
effective Lagrangian constructed for QCD as given in \cite{HZ,FHZ}. 
In Sect. 3 we demonstrate that it is theoretically possible to create 
$\thetaind \neq 0 $-states in heavy ion collisions.  The spectrum of the
Goldstone bosons produced in the nonzero $\theta$ background is calculated
in Sect. 4 for different geometries of the $\thetaind$-region. 
Finally, we end with concluding remarks and future considerations in Sect 5.

\section{Low Energy Effective QCD Lagrangian}

The starting point of our analysis is the low 
energy effective Lagrangian which 
reproduces the anomalous conformal and chiral Ward Identities.
The corresponding construction in the large $ N_c$ limit has been known 
for a long time\cite{Wit2,WV}. The generalization of the construction
of ref.\cite{Wit2,WV} for finite $N_c$ was given  in \cite{HZ,FHZ},
and we shall use formulae from the papers
\cite{HZ,FHZ}.  However, we should remark
at the very beginning   that all 
local properties of the effective 
Lagrangians for finite and
infinite $N_c$    are very much the same. Small quantitative differences in 
local physics between the description of \cite{Wit2,WV} 
on the one hand and 
the description of\cite{HZ,FHZ}
 on the other hand, do  not alter the qualitative 
results which follow.

The effective Lagrangian describes the light matter fields of QCD which 
consists of an octet of pseudo-Goldstone bosons ($\pi$'s, $K$'s, 
and the $\eta$), and the $\eta'$ singlet field. 
We parameterize these fields by a unitary matrix $U_{ij}$ corresponding to
the $\gamma_5$ phases of the chiral condensate 
($\la \overline{\Psi^i_R} \Psi^j_L \ra
	=-| \la \overline{\Psi_R} \Psi_L \ra|U_{ij}$)
in the following way: 
\beq
U=U_0\exp \left[ i\sqrt{2}\frac{\pi^a \lambda^a}{f_{\pi}}+
	i \frac{2}{\sqrt{N_f}} \frac{\eta'}{f_{\eta'}} \right],
\eeq
where $U_0$ solves the minimization equation for the effective potential, 
$\pi^a$ represents the pseudoscalar octet, and $N_f$ is the number of flavors.
The effective potential derived in \cite{HZ,FHZ} takes the following form:  
\beq
V(U,\theta)=-E\cos\lbrack\frac{1}{N_c}(\theta-i\log DetU)\rbrack
	-\frac{1}{2}Tr(MU+M^{\dagger}U^{\dagger}),
\label{poten}
\eeq
where $M$ is the diagonal quark mass matrix defined with their condensates: 
$M=-diag(m_i \la \overline{\Psi_i} \Psi_i \ra)$ and 
$E=\la b \alpha_s/(32\pi)G^2\ra$ 
is the vacuum energy with $b=11N_c/3-2N_f/3$. 
Expanding the cosine (this corresponds to the expansion
in $   1/N_c $) we recover the result of \cite{Wit2} at lowest order 
in $ 1/N_c $  together with the constant term $E$ required by the conformal 
anomaly:
\beq
\label{VVW}
V( U, \theta, N_c\rightarrow\infty) = - E - \frac{ \la \nu^2 \ra_{YM} }{2}
( \theta - i \log Det \, U )^2 - \frac{1}{2} \, Tr \, (MU + M^{+}U^{+} ) 
+ \ldots \; .
\eeq 
Here we used the fact that in the  large 
$ N_c $ limit $  E(\frac{1}{N_c})^2 = - \la \nu^2 \ra_{YM} 
$ where $ \la \nu^2 \ra_{YM}< 0 $ 
is the topological susceptibility in pure YM theory.
Corrections in $ 1/N_c $ stemming from Eq.(\ref{VVW}) constitute a new 
result of ref.\cite{HZ}. In order to 
demonstrate that the effective potential shown above is of the correct
form, we can verify that the following three features are 
characteristic of Eq.(\ref{poten}):

\begin{enumerate}
\item Eq.(\ref{poten}) correctly reproduces the Witten-Di
Vecchia-Veneziano effective chiral Lagrangian
\cite{Wit2} in the large $ N_c $ limit; 
\item It reproduces the anomalous conformal and chiral Ward identities of
	$QCD$;
\item It reproduces the known dependence in $\theta$ ({\it ie.} $2\pi$
	periodicity of observables) 
\cite{Wit2}.  
\end{enumerate}

To study the vacuum properties of the $\theta$ vacua
it is convenient to choose a diagonal basis to parameterize the 
fields $U$ as $ U= diag( \exp i \phi_q ), q=u,d,s $ 
such that the potential (\ref{poten}) takes the form: 
\beq
V(\theta,\phi_i)=-E\cos(\frac{\theta-\sum\phi_i}{N_c})-\sum M_i\cos\phi_i,
\label{potential}
\eeq
where $M_i$ are the diagonal entries of the quark mass matrix 
which was introduced 
above. Notice that in Eq.(\ref{potential}) 
the $\theta$ parameter appears only in
the combination $\sum \phi_i -\theta$. The minimum of the potential is given 
by the solution to the following equations:
\beq
\sin (\frac{\theta - \sum \phi_i}{N_c})=\frac{N_c M_i}{E}\sin \phi_i.
\eeq
In the limit where all the quark masses are equal and $M_i \ll E$, the 
approximate solution is given by $\phi_i \sim \frac{\theta}{N_f}$. 
In particular, in this limit
and for $N_f=2$,   
 the vacuum energy $V_{vac}(\theta)$ can be approximated as follows:
\bea
V_{vac}(\theta)\simeq V_{vac}(\theta=0)-2m|\la \bar{q} q \ra |
(|\cos\frac{\theta}{2}|-1), \nonumber \\
\langle \theta | \frac{\alpha_{s}}{8 \pi } G \tilde{G}
| \theta \rangle =-\frac{\partial V_{vac}(\theta)}{\partial\theta}=
-m|\la 0| \bar{q} q |0 \ra |\sin\frac{\theta}{2},
\label{V_vac}
\eea
which was used in the Introduction for estimates.
From this we see that a $\thetaind$-state is degenerate with the 
$\theta =0$ state in the chiral limit where $m_q=0$, as expected.

\section{Induced $\theta$-Vacua}

It was recently argued \cite{KFZ} that it may be possible 
to create an arbitrary $\theta \neq 0$ state in heavy ion 
collisions. In this scenario, we propose that the $\thetaind$-state 
is separated from our world $(\theta=0)$ by a shell of debris
expanding out at the speed of light. This idea is similar to
the so called disoriented chiral 
condensate that has been extensively 
studied in the last ten years \cite{DCC}-\cite{Rajagopal}. 

We would like to stress the  
point that an induced $\theta$-parameter is very different from
$\theta^{fund}$ which is zero in our world and which cannot be changed. 
The simplest way to visualize $\thetaind$
is to assume that right after the QCD phase transition the flavor
singlet phase of the chiral condensate is non-zero in a
macroscopically large domain. This phase is identified with
$\thetaind$. This identification is a direct consequence of the
transformation properties of the fundamental QCD Lagrangian under
$U(1)_A$ rotations by which the chiral singlet phase can be rotated
away at the cost of introducing $\thetaind$. From now on we will
refer to a $\thetaind$-state as simply a $\theta$-state, omitting 
the ``ind'' label.

Numerical lattice calculations in the past have provided strong evidence
that QCD undergoes a phase transition at a temperature in the range
of $150-200~MeV$. It is believed that above this critical point 
that chiral symmetry is restored and/or deconfinement occurs 
(quark-gluon plasma). The idea of the Disoriented Chiral Condensate (DCC)
arises when we consider what happens immediately 
after the phase transition upon cooling within heavy ion collisions.
DCC refers to regions of space (interior) where the chiral condensate 
points in a different direction from that of the ground state (exterior), and
separated from the latter by a hot shell of debris. The difference in energy 
between the created state and the lowest energy state is proportional to the
small parameter $m_q$ (see Eq.(\ref{V_vac})) for both the DCC and
the $\theta$-state and is
therefore negligible at high temperature. 

First, we recall the DCC scenario as given in \cite{RW} with emphasis 
on the analogy between DCC and $\thetaind$-state. 
Rajagopal and Wilczek \cite{RW} use the $O(4)$ linear
sigma model to describe the low energy dynamics of the pions and the chiral
condensate. All fields are represented by a 4-vector with components 
$\phi=(\sigma,\vec{\pi})$, where $\sigma$ represents the chiral condensate 
and $\vec{\pi}$ represents the triplet of pions. Throughout the exterior  
region the vacuum expectation value of $\phi$ is $(v,0)$. In the interior 
region, however, the pion fields can become non-zero and $(\sigma,\vec{\pi})$
wanders in the four dimensional configuration space. The
high energy products (shell of debris) 
of the collision expand outwards at relativistic speeds and separates
the misaligned vacuum interior from the exterior region.

In \cite{RW} all calculations were done under the assumption
of a quenched system. In the quenched approximation, the  
$(\sigma,\vec{\pi})$ fields are suddenly removed from a heat bath 
$(T \geq T_c \sim 200 MeV)$ 
and evolved according to the zero temperature equations of motion. 
This is done numerically by giving the fields a non-zero 
vacuum expectation value $\la \phi \ra \neq 0$
and letting this field configuration evolve in time according 
to the zero temperature equations of motion. 
The hope is that regions of misaligned vacuum with an arbitrary
isospin direction will be created. In \cite{RW} it was shown that if the 
cooling process is very rapid and the system is initially
out of equilibrium, there
is a temporary growth of long wavelength spatial modes of the pion field
corresponding to domains where the chiral condensate is approximately
correlated. In the case of DCC, the created
state will relax to the true vacuum by coherent emission of pions 
with the same isospin orientation producing clusters of charged or neutral 
pions. 
To be more specific, let us consider the case $N_f=2$. The matrix 
$U$ is parameterized by the misalignment angle $\phi$ and the
unit vector $\vec{n}$ in the isospin space:
$$ U=e^{i\phi(\vec{n}\vec{\tau})} \; , U
U^{+} = 1 \; , \la \bar{\Psi}_{L}^{i}
\Psi_{R}^{j} \ra 
=  - | \la \bar{\Psi}_{L} \Psi_{R} \ra | \, U_{ij}.
$$
The energy density of the DCC is determined by the mass term:
\beq
\label{2}
 E_{\phi}= -\frac{1}{2} Tr( M U + M ^{\dagger}U^{\dagger}) = 
 - 2m  | \la \bar{\Psi}  \Psi  \ra | \cos(\phi).
\eeq
Eq. (\ref{2}) implies that that any $\phi\neq 0$ is not a stable
vacuum state because $\frac{dE}{d\phi}\neq0$, i.e. the vacuum is misaligned. 
Since the energy difference between the misaligned state and the
true vacuum is proportional to $m_q$, the probability to create
a state with an arbitrary $\phi$ at high temperature $T \sim T_c$ is 
proportional to $\exp [V(E_{\theta}-E_0)/T]$ and depends on 
$\phi$ only very weakly, i.e.
$\phi$ is a quasi-flat direction. Right after the phase transition when 
$\la\overline{\Psi}\Psi\ra$ becomes non-zero, the pion field oscillates.

Now we turn to our main subject, the $\theta$-vacuum state. 
First we would like to comment that even though the 
$\theta$ parameter is related to 
the large constant $E$ in Eq. (\ref{potential}), $\theta$ dependence is 
actually proportional to $m_q$ since in the 
chiral limit all dependence can be removed by performing 
a $U(1)_A$ rotation of the chiral condensate phases. When the combination
of $(\sum\phi_i-\theta)$ is close by an amount $O(m_q)$ to its 
vacuum value, the Boltzmann suppression due to the term $E$ at high
temperature is absent. This is essentially what allows the induced 
$\theta$-state to be formed.
  
To model the creation of an arbitrary $\theta$-state, we will use the
effective chiral Lagrangian   described in the previous Sect. 2. 
We consider three flavors of massive quarks, and derive the following 
equation of motions:

\beq
\label{eom}
\ddot{\phi_i} + \nabla^2 \phi_i 
  + E\sin(\frac{- \theta + \sum \phi_i}{N_c}) 
 - M_i \sin(\phi_i) - \gamma \dot{\phi_i} =0,\:\:i=u,d,s,
\eeq
where the constant $\gamma$ in  front of $\dot{\phi}$ represents
the damping of the system which can be attributed to various sources
such as expansion or the emission of pions. 

In order to model the situation numerically, several 
assumptions are made. First, assume that the $\theta^{ind}$ parameter acquires 
a non-zero value when the temperature T $\gg m_q$ and the system 
is out-of-equilibrium. Once again, $\theta^{ind} \neq 0$ follows from the 
assumption that the mean value of singlet phase $ \la \sum \phi_i \ra $ 
is not zero over a macroscopically large domain immediately
upon cooling through the phase transition. 
Performing a $U(1)_A$ rotation then gives us 
an equivalent description  with $\theta^{ind} \neq 0$. 
The second assumption is that the phases $\phi_u$, $ \phi_d$, and
$\phi_s$ have small random values (modulo $\pi /3$) at these high 
temperatures. The final assumption is that the cooling process is 
very rapid and the 
system obeys the zero temperature equations of motion after this
initial quench. There is obviously some sort of damping mechanism 
present in any real system, and this damping can be attributed to 
the expansion of the region inside the shell and/or the radiation of
Goldstone bosons. The damping constant, $\gamma$, 
was chosen to be of the same order order 
of magnitude as $\Lambda_{QCD}$, which should be a  
reasonable value according to all other scales in the system.
	
The numerical algorithm used to solve these coupled differential 
equations was a fourth-order Adams-Bashforth-Moulton predictor-corrector 
method. The phases were initially assigned random values according to a 
uniform distribution ($|\phi_i| < \pi /3$). The phases
were rescaled in such a way as to be dimensionless. We evolved 
the equations for 6000 time steps of size $10^{-27}~sec$ on 
a cubic lattice with spacing $a=0.005~MeV^{-1}=1~fm$. In order to 
examine the momentum dependence, the Fourier transform \cite{fourier}
of the field configuration was calculated at periodic intervals 
in time. The data was then binned according to the magnitude
of the wave vector $\vec{k}$ so that the dependence on $|\vec{k}|$ 
could be examined. 
We consider a volume of $(8~fm)^3$. In all calculations that follow, we
use the current quark masses $m_u=4~MeV, m_d=8~MeV,$ and $m_s=150~MeV$.

\begin{figure}
\epsfysize=3.5in
\epsfbox[ 17 544 331 741]{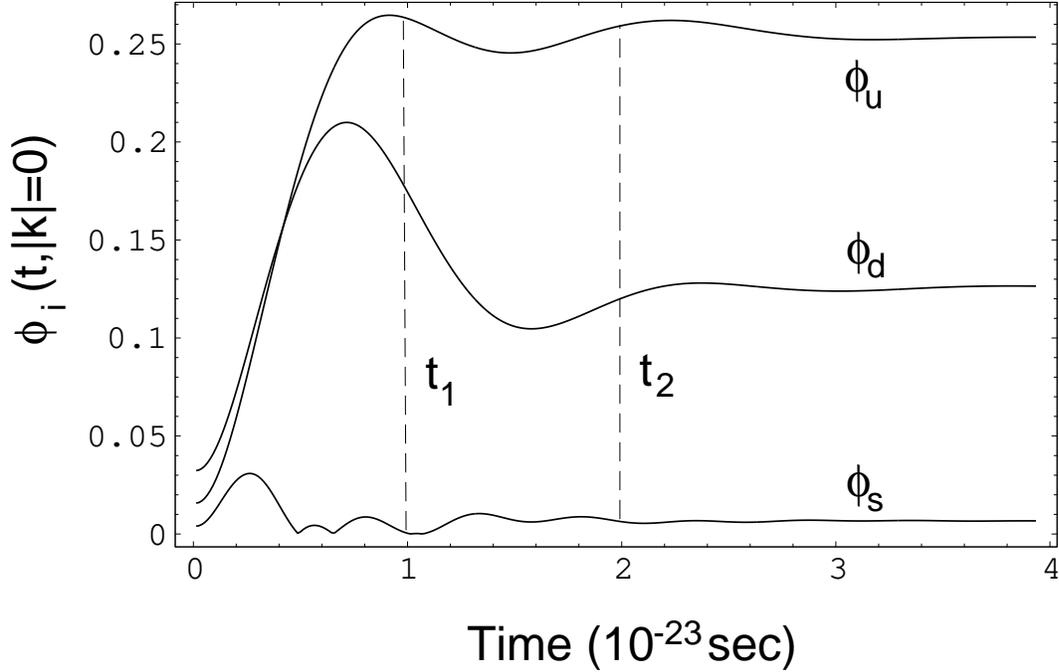}
\caption{$|\phi_i(k=0)|$ is plotted as a function of time for the up, down, 
and strange quark. Notice that the zero momentum modes 
of the $\phi_i$ fields settle to a non-zero value in a time on the
order of $10^{-23}~s$. The times $t_1$ and $t_2$ represent the values we 
chose for $\tau_{shell}$, the the time when the shell separating
the two regions disappears. }
\label{phis}
\end{figure}

In Fig. \ref{phis} we show that the zero modes of the $\phi_i$ fields 
settle down to a constant value in a time $\tau \sim 10^{-23} s$. 
In the same time interval, all other modes die off. This, along with 
a test of volume independence, shows that a true non-perturbative 
condensate has been formed. 
In the case where all quark masses are equal, the 
zero mode is expected to settle down to $\phi_i \sim \theta/N_f$, 
which has also been verified numerically. 

In order to incorporate the effects of expansion, we 
could introduce a time dependent lattice spacing $a(t)$. If $a(t)$ increases 
with time, this will model the dilution of energy in the system. Following 
an analogy with an expanding universe, the coefficient of the damping term
$\dot{\phi_i}$, $\gamma$ goes like $constant\times\dot{a}/a$. Although we
have not taken this into account in our analysis, we can 
estimate the result as follows.
The Laplacian term in Eq.(\ref{eom}) is 
numerically proportional to $1/a$, and if $a(t)$ increases with time, the 
system will require more time to relax to the $\theta$-state. If the
coefficient, $\gamma$, is altered as mentioned above and $a(t)$ increases
at a constant rate, the damping term will become smaller as time 
goes on. The overall effect of a time dependent lattice spacing is to 
increase the formation time of the $\theta$-state. 

\section{Goldstone Bosons}
Now that we have argued the induced $\theta$-state may be formed in
heavy ion collisions we would like to discuss how to detect them
experimentally.   
Naturally, the first thing to look for would be axion production.  
If axions\cite{axrev} do exist, they  can be produced during the 
relaxation of  $\thetaind \neq 0$
to the lowest energy state with $\theta= 0$.  
However, the axion production rate under the conditions which
can be achieved at RHIC will be too low in comparison with  the limit 
already achieved from the astrophysical and cosmological
 considerations\cite{axrev}. Therefore, the question remains:
``How could the the formation of a $\theta$-state be detected
experimentally?'' Fortunately there are a number of ways in
which this might be done.  One proposed possible signature is related
to the observation of specific $CP$ odd correlations of the Goldstone bosons
\cite{Pisarski} but the effect would probably be washed out by final state
interactions.  A second possible signature was proposed in \cite{arz}
which comes about because of the unusual 
properties of the Goldstone bosons and  $\eta'$ 
in the $\theta$-background.  In the
$\theta$-world the masses of these particles are altered and for
example the mass of $\pi^0$ would be noticeably smaller.  As well,
these particles become a mixture of pseudoscalar/scalar rather than pure
pseudoscalars as for $\theta=0$.  This fact means that the $CP$ odd
decays $\eta\rightarrow\pi\pi$ and $\eta'\rightarrow\pi\pi$ are no
longer suppressed in the $\theta$ background and become of order 1
\cite{arz}. The full width of the $\eta',\eta$ decays in our world are much
smaller than the above decays in the $\theta$-world:
\bea
\Gamma_{\theta \neq 0}(\eta'\rightarrow \pi\pi)\sim 2~MeV \gg \Gamma_{\theta =0}^{total}(\eta')\sim 0.2~MeV \: (experimental), \\
\Gamma_{\theta \neq 0}(\eta\rightarrow \pi\pi)\sim 0.2~MeV \gg \Gamma_{\theta=0}^{total}(\eta)\sim 118~keV \: (experimental).
\eea
Thus observation of an increased rate of decay of $\eta',\eta$ at masses 
slightly shifted from the accepted values would be a signature of the
$\theta$-vacuum.  However, this signal may be a difficult one to 
distinguish in the large amounts of data generated in heavy ion
collisions.  Therefore we would like to discuss a signature that would
be easier to experimentally observe. 
In this work we consider an additional new signature which 
could possibly verify if a
$\theta$-state is created in heavy ion collisions. As was discussed
briefly in the introduction, the creation of a $\theta$-state
could greatly enhance the production of low momentum $(\sim 10~MeV-100~MeV)$
Goldstone bosons.

Before going into the details we would like to
discuss the concepts behind this idea.  We have already presented some arguments 
that a $\theta$-state may be formed in heavy ion collisions if
protected from the exterior $\theta=0$ world by the shell of debris.
At some point, however, this shell ceases to isolate the interior and
the influence of the exterior world will be felt.  The way in which
this happens is not well understood, but we would like to use 
an instantaneous approximation in order to obtain some
feeling about the possible outcome. There are
basically two time scales present in the system: 
the lifetime $\tau_{shell}$ of the shell 
 separating the two worlds and the scale
associated with strongly interacting particles,
$(\Lambda_{QCD})^{-1}$. If these differ by orders of magnitude, the
disappearance of the shell can be considered either as an
instantaneous perturbation or an adiabatic perturbation.  

Realistically, we expect $\tau_{shell}$ to be about the same order of
magnitude as $(\Lambda_{QCD})^{-1}$ and therefore the disappearance of
the shell is somewhere between an instantaneous process and an
adiabatic one.  However, we do not yet know how to treat the process
properly so we will use the instantaneous approximation in order to
obtain an order of magnitude estimate of the spectrum of emitted particles.

The instantaneous perturbation is an approximation that is well known
in quantum mechanics. For our specific problem 
we essentially   assume that
all states which were formed in the $\thetaind$ 
background will suddenly find themselves  in the new vacuum state
 (with $\theta=0$) after the shell
{\it instantaneously} blows apart.   These states have no choice but to
transform to 
the asymptotic states of this new (for them) $\theta=0$ world. 
The  dynamics of this strong transformation is quite complicated 
and beyond the scope of  the present work. However, if this transition is
instantaneous, the procedure to do the calculations for
such a transition  is well known: we assume that 
the initial state has no time to adjust its properties
when the shell decays, and therefore, we expand the initial state
in terms of the asymptotic states of $\theta=0$ world.
 
In more detail, we do the following.  The
field values, $\phi_i$, obtained for the $\theta$-state as in previous
work \cite{KFZ} are embedded into a larger grid where the field values
take their $\theta=0$ vacuum values. This data now contains a plateau
of field values corresponding to $\theta$-state values surrounded
by zero field values.  This field configuration must now resolve itself 
into asymptotic free particles.  With this process in mind we determine
the momentum spectrum of free particles corresponding to this distribution
by considering the Goldstone boson fields as true quantum fields at this 
instant. 

In order to calculate the spectrum, we must obtain the 
quantities $a(\vec{k})$ and $a^{\dagger}(\vec{k})$. 
These are obtained by performing the Fourier transform of the field 
configuration at time $t=\tau_{shell}$ when the shell breaks down,
\beq 
a(\vec{k})=\int d^{3}\vec{x}\psi(\vec{x})\exp(-i\vec{k}\cdot\vec{x}),
\label{ak}
\eeq
where $\psi(\vec{x})$ is the distribution amplitude of the $\phi_i$ fields
obtained from evolving the equations of motion (Eq.(\ref{eom})) on a cubic
lattice, as was done in \cite{KFZ}.  
From Eq.(\ref{ak}),  the number operator is given by 
$N(\vec{k})=N_{0}a(\vec{k})a^{\dagger}(\vec{k})$ where $N_{0}$ is an 
overall constant that will be determined using the following argument. 
As we mentioned earlier, the $\theta$-state differs from the true vacuum 
state by a small amount of energy $\sim m_q$.
The amount of energy that is available when the $\theta$-vacua decays
is obtained by analyzing the $\theta$ dependence of the vacuum energy 
density $E$:
\beq
E_{\theta}=m_q|\la \overline{\Psi}\Psi \ra| N_f \cos(\frac{\theta}{N_f}).
\eeq
Therefore, the amount of energy $\Delta E ~ V$ 
available due to the formation
of the $\theta$-state as discussed above  is given by
\beq
\Delta E ~V = (E_{\theta=0}-E_{\theta}) \cdot V
  \simeq 20 \left( \frac{V}{(10~fm)^3} \right) ~GeV,
\label{deltaE}
\eeq
where $V$ is the the volume of the created $\theta$-region. 
If we take for example a $\theta$-state with
volume $V=(8fm)^3$ the amount of energy available is 
$\Delta E ~V=10~GeV$. This 
represents only a small amount of energy compared to the $cm$ energy,
$\sqrt{s}=40~TeV$, expected in heavy ion collisions.
If the total amount of available energy is known, the constant $N_0$
can be fixed by enforcing the following conservation of energy
constraint 
\beq
\Delta E ~V=\sum_{i}\int \frac{d^{3}\vec{k}}{2(2\pi)^3}
N_{0}a_i(\vec{k}) a^{\dagger}_i(\vec{k}),
\eeq
where the sum is performed over all types of particles considered
({\it ie.} all neutral Goldstone bosons). Once this constant $N_0$ is fixed,
the total number of pions and etas can be calculated by 
\beq
N=\int\frac{d^{3}\vec{k}}{2\omega_k(2\pi)^3}
N_{0}a(\vec{k})a^{\dagger}(\vec{k}),
\label{number}
\eeq
for the simplified case of only one collision of two   nuclei. If 
we consider that RHIC will collide gold nuclei at a rate of central collisions
of about 1 $kHz$, we can estimate the total number of these low 
momentum particles produced by simple multiplication. In what follows 
we present all results for just one event.

As we have already described, we proceed as follows.  
At some time $t$ before the
fields $\phi_i$ settle down to the constant field configuration (see
Fig. \ref{phis}), we take the position space data and embed this in a
larger square grid where the field values are zero (our world). 
We assume an instantaneous perturbation and then
take the fourier transform of this field configuration in order to
obtain the operators $a(\vec{k})$ and $a^{\dagger}(\vec{k})$. It may
be argued that the instantaneous approximation is not completely justified
 in this case as we expect $\tau_{shell} \sim \tau_{QCD}$ and therefore this
calculation gives us a rough estimate at best.
 
\begin{figure}
\epsfysize=3.15in
\epsfbox[ 15 559 335 751]{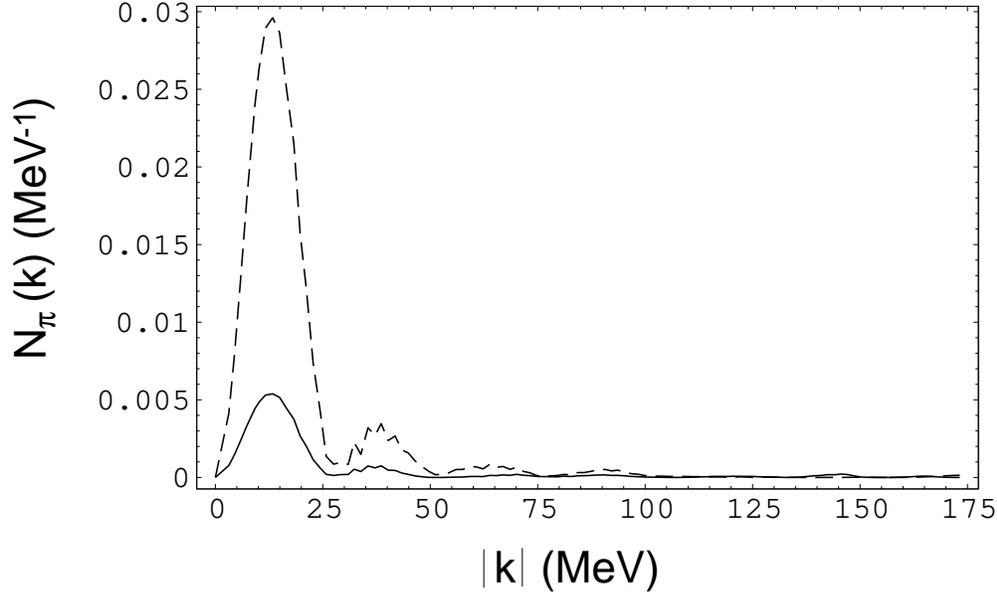}
\caption{The number of neutral pions produced, $N_{\pi^0}(k)$, is plotted 
as a function of the magnitude of the wave vector, $|\vec{k}|$. 
Above we show that the momentum distribution of the $\pi^0$-mesons 
produced is primarily $< 25~MeV$ for two different values of $\tau_{shell}$. 
The solid line represents the earlier time $t_1$ and the dotted line represents
the later time $t_2$ (see Fig. \ref{phis} for the positions of $t_1$ and 
$t_2$ in the evolution of the $\phi_i$ fields). 
This is expected as the formation of the $\theta$-state 
can be attributed to the enhancement of the low momentum modes.}
\label{sqpizero}
\end{figure}

In order to analyze the spectrum of the diagonal (neutral) components 
of the matrix $U$ ($\pi^0,\eta,\eta'$) we must analyze the following 
combination of the $\phi_{i}'s$:
\bea
\pi^0=\frac{f_{\pi}}{2\sqrt{2}} (\phi_u -\phi_d), \nonumber \\
\eta=\frac{f_{\pi}}{2\sqrt{6}} (\phi_u +\phi_d -2\phi_s), \\
\eta'=\frac{f_{\eta'}}{2\sqrt{3}} (\phi_u+\phi_d+\phi_s) \nonumber .
\eea
For all numerical calculations, we can take the constant $f_{\eta'}$ to be
$\sim f_{\pi}$. One of the crucial decisions which must be made is when
to choose the time at which the shell breaks down. We chose several values 
for $\tau_{shell}$. 
The majority of the 
particles which are produced due to the formation of the 
$|\theta \ra$-state have momentum $k < 25 MeV$
( for our choice of the volume $V=(8fm)^3$). In Fig. \ref{sqpizero} 
we show a plot of $N_{\pi^0}(k)$ as a function of $|\vec{k}|$ 
for the neutral pion. The function $N_{\pi^0}(k)$ has dimensions $MeV^{-1}$,
so that the total number of particles, $N_{\pi^0}$, is a dimensionless number
given by:
\bea
N_{\pi^0}&=&\int^{\infty}_{-\infty} dk N_{\pi^0}(k) \nonumber \\
       &=&\int^{\infty}_{-\infty} dk \frac{4\pi}{2(2\pi)^3w_k}
N_{0}a(\vec{k})a^{\dagger}(\vec{k}).
\eea
The two different lines represent the times at which we assumed the shell
to break down. The solid line represents the earlier 
time ($\tau_{shell}=t_1=1500/6000$ time steps) while the dotted line 
represents  $\tau_{shell}=t_2=3000/6000$ time steps (this representation
for $t_1$ and $t_2$ will be followed in all graphs that follow).
For this calculation, we assumed the
created $\theta$-state had a volume of $(8~fm)^3$   which exists in a 
larger volume of $(64~fm)^3$. The explanation of this phenomenon is simple:
at time $t_1$ the amplification of the zero mode is not as profound  as at
time $t_2$. Furthermore, at time $t_1$ a considerable portion of the 
energy goes to high momentum particles.

Below we show the same graph for the $\eta$ and the $\eta'$. 
Notice that that the $\eta'$ is produced in large amounts compared 
to the $\pi^0$-mesons and the $\eta$-mesons. In order to obtain the 
total number of
each particle produced, we use Eq.(\ref{number}) to calculate the total
number of particles produced per collision: 
$N_{\pi^0}=0.1$, $N_{\eta}=2.8$, $N_{\eta'}=19.3$ for
$t=t_1$ and $N_{\pi^0}=0.5$, $N_{\eta}=4.3$, $N_{\eta'}=18.4$ for
$t=t_2$.
We have also checked that in the limit 
where all quark masses are equal, only the $\eta'$ is
produced, as expected.  
It should be noted that, as demonstrated by these figures and the 
coarsening phenomenon (i.e. the phenomenon
of amplification of the zero mode as time increases), we would expect that
as $\tau_{shell}$ increases up to a maximum value, 
the majority of the particles would reside in the low momentum regime. 
We consider our estimate as  a very conservative one  due to the fact 
that we do not count  the particles  $\eta$ and   $\eta'$
which will be produced during the formation period  of the 
$\theta^{ind}$ state. These particles will have higher momentum modes 
and can not be easily distinguished from the large number of the 
similar particles from the background.
Therefore, the main conclusion of our analysis is as follows:  
if  induced $\theta$ state is produced than it  will result in the
strong enhancement of emission of  low-energy $\eta$ and $\eta'$ 
mesons which clearly signals the existence of the induced $\theta$ state.
\begin{figure}
\epsfysize=3.15in
\epsfbox[ 14 561 331 751]{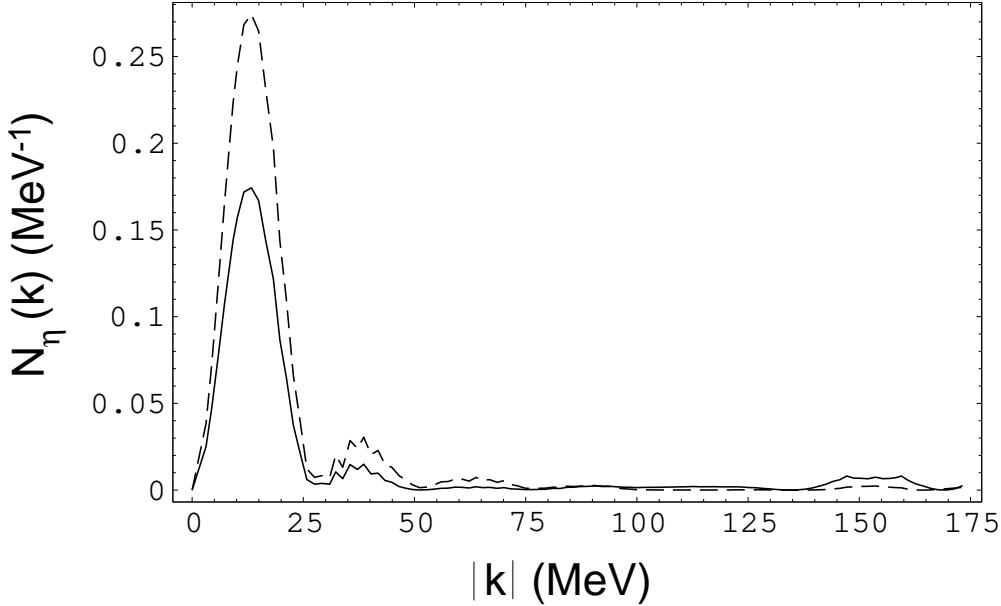}
\caption{$N_{\eta}(K)$ is plotted as a function $|\vec{k}|$, 
at two different times $t_1$ and $t_2$ as in 
Fig. \ref{sqpizero}. The higher peak represents the
later time with the  main difference being 
that now a larger percentage of the produced particles lie in 
the $k<25~MeV$ range.}
\label{sqeta}
\end{figure}
\begin{figure}
\epsfysize=3.15in
\epsfbox[ 10 553 335 750]{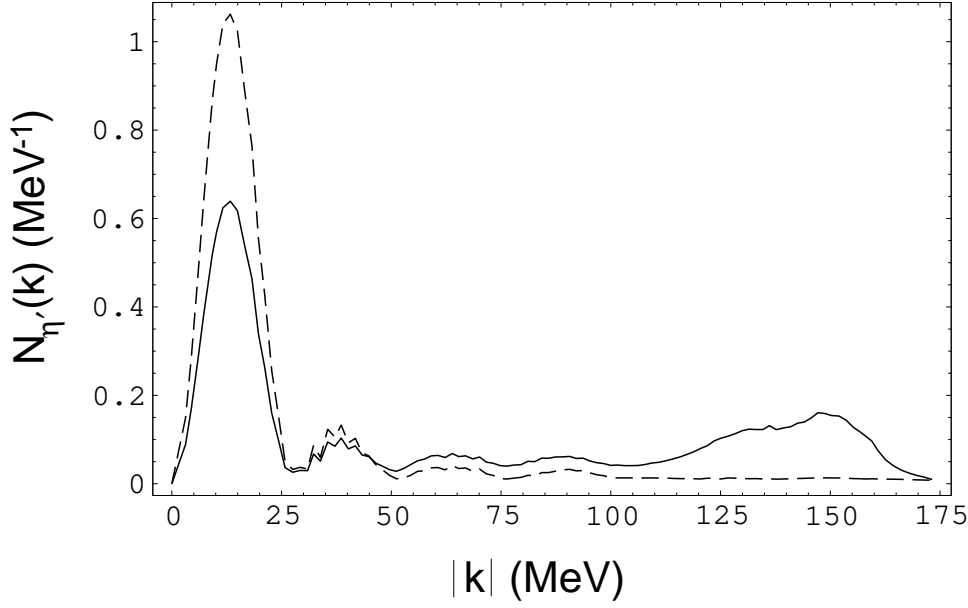}
\caption{$N_{\eta'}(k)$ vs. $|\vec{k}|$ is shown above. 
This graph shows that the $\eta'$ clearly dominates
the spectrum compared to the other neutral particles.}
\label{sqeta'}
\end{figure}

When our analysis was completed, we became aware that a
similar conclusion has been reached  in 
a recent paper by Baier {\it et al.} \cite{eta'prod} where    the production
of an excess of $\eta'$-mesons is discussed 
in the scenario suggested in \cite{Pisarski}.
Despite the fact that the starting points of the present work  and  
\cite{eta'prod}
are quite different (we start from  the induced $\theta$ state while
in paper  \cite{eta'prod} the starting point
is a metastable state which acts like a  region with a non-vanishing
$\theta$ angle), the general conclusion is very similar. Namely, 
a strong enhancement of emission of low-energy $\eta'$-mesons 
is expected irrespectively of the origin of the initial unusual state.   
The details, however, are quite different. 
The main difference is the following.
We assume that a $\theta$-state is produced and 
instantaneously decays such that   we calculate the particle production 
due to this single event, not taking into account previous particle 
production during the formation period. 
As was mentioned above, we expect that 
those  particles will have much higher momentum modes and we ignore them. 
The approach of \cite{eta'prod}
is different from ours as their calculation of the 
number of produced $\eta'$-mesons  is based on the specific
mechanism of the decay of the original metastable state
\cite{Pisarski}. Such a decay receives contributions from the whole 
evolution, not just one particular instant. 
This difference explains the difference
in rates: for a domain with radius 5 $fm$, they estimate that about 90-100 
$\eta'$-mesons would be produced while our estimation is $4-5$ 
times smaller.

In the above calculations of $N(\vec{k})$ the geometry of the created 
$\theta$-state is assumed to be a cubical box. 
For the case of heavy ion collisions, the more
realistic case is that of an ellipsoid or elongated rectangular box. 
Upon collision, we expect
the collision to create a region of quark-gluon plasma which
has an asymmetry in one direction. In order to model 
this situation, we evolve a rectangular grid and embed it in a larger square 
grid and compute the Fourier transform. Once again, we consider the
angular averaged value of $N(\vec{k})$.
\begin{figure}
\epsfysize=3.15in
\epsfbox[ 16 549 330 749]{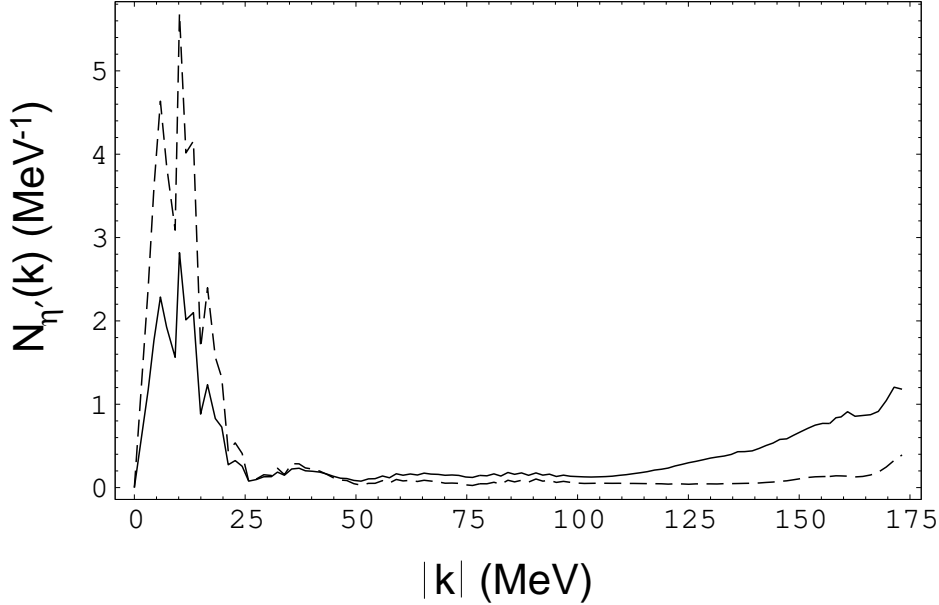}
\caption{ The momentum distribution ($N_{\eta'}(k)$vs.$|\vec{k}|$) 
is now shown for the case where the geometry of the $\theta$-region
  is considered as a rectangle. We consider the case where the size of
the $\theta$-region is larger along the longitudinal directions
compared to the transverse directions. The solid and dotted lines once again
represent times $\tau_{shell}=t_1$ and $t_2$ respectively. Although the 
majority of the spectrum still exists in the low momentum range, 
the higher momentum modes now become more evident.}
\label{receta'}
\end{figure}

For a $\theta$-region with 
dimensions $32~fm \times (8~fm)^2$, we show the spectrum of the 
$\eta'$ at the same times as shown in Fig. \ref{sqeta'}. The side of 
length $32~fm$ is the direction parallel to the beam direction.
The low momentum modes still dominate, but the peak is not as 
sharp and the higher modes show a stronger presence. 
Also, since the $\theta$-region is now 
larger, from Eq.(\ref{deltaE}) we would expect more particles to 
be created since there is now more energy available. This is evident 
when comparing Fig. \ref{sqeta'} with Fig. \ref{receta'}. For the two 
different times shown, we find that $N_{\pi^0}=0.3$, $N_{\eta}=12.7$, 
$N_{\eta'}=76.3$ for
$t=t_1$ and $N_{\pi^0}=1.7$, $N_{\eta}=16.8$, $N_{\eta'}=73.8$ for
$t=t_2$.

\begin{figure}
\epsfysize=3.15in
\epsfbox[ 9 503 344 718]{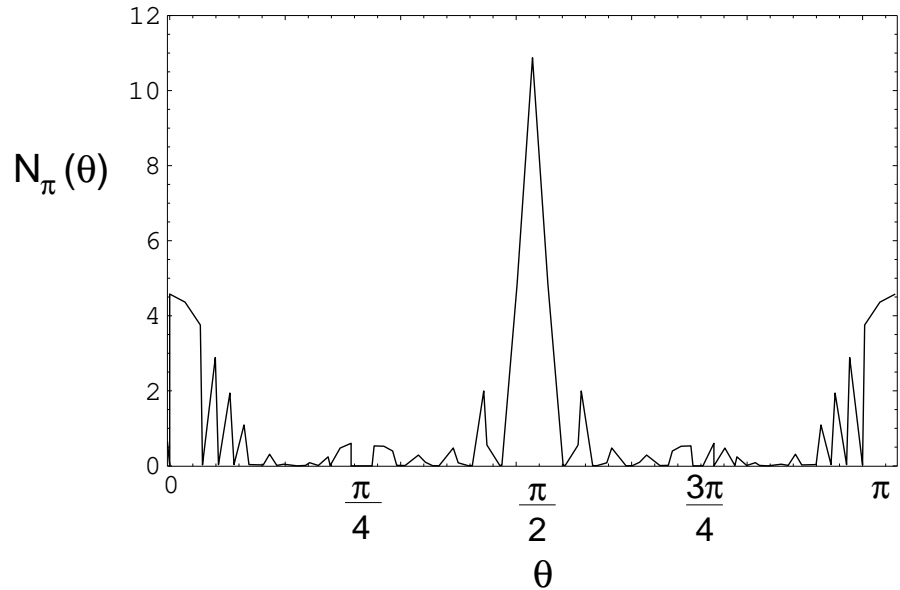}
\caption{ For the scenario shown in Fig. \ref{receta'}, we consider the 
angular dependence on the azimuthal angle for the $\pi^0$ instead 
of the $\eta'$. The azimuthal angle is defined so that $\theta=0$ coincides
with the beam axis. }
\label{pizerothetaangdist}
\end{figure}

In this case where spherical symmetry is lost, we must 
also examine the angular distribution of the system. 
We choose the coordinates such that the long edge of the 
$\theta$-region is parallel to the z-axis.
In Fig. \ref{pizerothetaangdist}, we show the angular distribution of
$N_{\pi^0}(\theta)$ as a function of the azimuthal angle $\theta$. This plot 
suggests that the majority of the particles would have larger 
momentum components in the direction perpendicular to the beam direction.

As was shown in Figs. \ref{sqpizero},\ref{sqeta},\ref{sqeta'}, and 
\ref{receta'}, if a $\theta$-state can be created in heavy ion collisions
we should expect an excess of low momentum Goldstone bosons.
In particular, the $\eta'$ would be produced in large amounts.
Besides that,  
the  low momentum $\pi^0$, $\eta$, and $\eta'$-mesons would considerably  
increase the photon and $e^+e^-$ pair production through such decays as 
$\pi^0\rightarrow \gamma e^+ e^-$, $\eta \rightarrow \gamma e^+ e^-$, and
$\eta' \rightarrow \gamma e^+ e^-$. 
In particular, it would result in  enhancement of low-energy  dileptons, 
which could possibly provide a solution to the observation of an excess 
of these particles seen at CERN \cite{cern}. Indeed, as was demonstrated 
by Kapusta {\it et al.} in \cite{kapusta} a sufficiently large 
enhancement\footnote{The enhancement mechanism of $\eta'$-production 
suggested in \cite{kapusta} is quite different from what we have considered
in this paper. However, the general conclusion (that $\eta'$ enhancement
leads to the excess of dileptons) is insensitive to the nature of the 
$\eta'$-mesons. Only by performing a detailed experimental analysis of 
the momentum and angular distributions can one mechanism be chosen over 
another.} of $\eta'$-production could easily explain the excess of 
dileptons seen at CERN \cite{cern}. The large amount of produced 
$\eta'$-mesons in our work is a direct consequence of the decay of the 
induced $\theta$-state. 

\section{Conclusion}

In this work we showed that the consequences of creating non-trivial
$\theta$-vacua in heavy ion collisions could be the 
amplification of production of light Goldstone bosons in the $10~MeV$ momentum 
range. In all calculations, we worked in the instantaneous 
approximation where the shell separating the $\theta$-state 
disappears in a time much less than any internal time scale. 
These low momentum particles could possibly then decay to 
low momentum photons and dileptons,
which could be easily detected. We would like to make the 
suggestion that this could possibly account for the unexplained
abundance of low momentum dileptons observed at CERN
\cite{cern}. We are thankful to Dima Kharzeev for correspondence regarding 
the relevant work in \cite{kapusta}.


\end{document}